# PROTECTOR CONTROL PC-AODV-BH IN THE AD HOC NETWORKS

S. Boujaada[1], Y. Qaraai[1], S. Agoujil[1] and M. Hajar[2]

E3MI[1] and ROI[2] teams,Computer Science[1] and Mathematics[2] ,
Departments Faculty of Sciences and Technologies, B.P. 509 Boutalamine 52000
Errachidia, Morocco.

## ABSTRACT

*In this paper we deal with the protector control that which we used to secure AODV routing protocol in Ad Hoc networks. The considered system can be vulnerable to several attacks because of mobility and absence of infrastructure. While the disturbance is assumed to be of the black hole type, we purpose a control named "PC-AODV-BH" in order to neutralize the effects of malicious nodes. Such a protocol is obtained by coupling hash functions, digital signatures and fidelity concept. An implementation under NS2 simulator will be given to compare our proposed approach with SAODV protocol, basing on three performance metrics and taking into account the number of black hole malicious nodes.*

## 1. INTRODUCTION

In the wired networks, the computers are connected by broadcast cables and they are characterized by their powers in terms of capacity for treatment and storage. Moreover, such networks offer a stable band-width, a good and inexpensive quality [11]. In the early 90's the consequent evolution carried in the wireless networks made the interest of the mobile computing grow. The latter offers a flexible mechanism of communication enters the users and an access to all services available in a typical environment (fixed) through an independent network of the physical location (geographic) and user mobility [11].

The most usual wireless networks deployed today are based on fixed infrastructures: sites accommodating the base stations in the case of the cellular networks or cables for wired infrastructure. A connectivity between the various elements in the network is organized and centralized [11].

The Ad Hoc are wireless networks capable to be organized without infrastructure previously defined. For example from one device to another without any infrastructure [7]. Each node in the network is equipped with a radio interface and it is free to join, leave and move independently. As a result, the network topology changes rapidly. To meet the need, the network may change rapidly and spontaneously and configures in an autonomous way according to the existing connections between nodes. In the Ad Hoc networks, node should have the capability to function in the same time routers and terminals. Moreover, the communication between nodes is ensured dynamically [9].

When the Ad Hoc nodes communicate, it is interesting to define a good strategy for transferring data by taking into account at any time the characteristics of the network such as the dynamic topology, number of links, band-width, and network resources [10]. Moreover, the method adopted must offer the best routing of data. The routing protocol in a mobile Ad Hoc network can be categorized into two classes: proactive (definition of the routes in advance) and reactive (it is with the request that the route will be defined) [9].

The concern of the security in the routing operations represents a principal challenge in the design of the routing protocols. Indeed, due to lack of such infrastructure or assumption of central administration, in contrast the traditional security solutions are not adapted to cope with the features of the Ad Hoc networks. Several vulnerabilities exist in these networks: manufacturing, modification, selfish or malicious nodes, usurpation of identity or suppression of the traffic in the network, the black hole attack [20], the worm hole attack [13], and so on.

Each node in the network contributes to the good performance in the routing; in contrast each element represents a point of vulnerability. In particular, if no mechanism is set up to make it possible for each node to determine the good performance and to check the coherence of the routing data, the node accepts the information of routing coming from any other node in the network. That is an attacker can send messages containing incorrect information on the network, in order to conduct a malicious action.

For this reason, the traditional mechanisms of security and the protocols are not directly applicable and require a suitable securing in the Ad hoc networks. Several researches explored a variety of mechanisms to answer the problems of data security, and a certain number of secure routing protocols have been suggested in order to prevent different types of attacks (TAODV, ARAN and SAODV [19, 21, 22]).

In this work, which is part of the Ad Hoc networks, the routing is assumed to be provided by using reactive routing protocol AODV. In this case, we talk about a system in the autonomous case. When the system is subject to a disturbance, as the case of the black hole type, it may be that it prevents the good routing performance because of the presence of the malicious nodes. Thus, we are interested with the problem of controlling the data routed through AODV in the presence of such a disturbance. This is the protector control of AODV protocol facing the black hole attack: PC-AODV-BH. This control is achieved by a coupling of the security mechanisms (digital signatures, hash functions) with the concept of fidelity associated with the nodes in the considered network.

The principle of protector control consists (Qaraai et al. [16]), in the case of AODV protocol, to develop and implement algorithms making it possible to return the disturbed system in its autonomous state, while trying to neutralize the effects of black hole attacks. Hence, the terminology of the resulting protocol PC-AODV-BH.

The organization of this paper is as follows: In section 2 we give an overview of the AODV routing protocol and the black hole attack. After dealing with the history of some protocols introduced to secure AODV protocol, we give a detailed approach followed by the proposed algorithms in the third section. The last section will be devoted to the simulation tests by considering some metrics, while varying the number of black hole attacks. A comparison of PC-AODV-BH and SAODV protocols will be made to test the effectiveness of introduced control.

## 2. ROUTING AND ATTACKS

This section provides a brief recall on the reactive AODV routing protocol and also a preview of the black hole attacks. In other words, it's about the principal elements defining respectively the autonomous and the disturbed systems.

### 2.1 AODV PROTOCOL

The AODV was developed by Charles E. Perkins and Elizabeth M. Royer [14]. It is a reactive protocol based on the principle of distance vector. This protocol uses two mechanisms named route discovery and route maintenance. Besides the routing node by node, it builds the roads by using a cycle of requests called route request and route reply (RREQ and RREP) [4]. It uses destination sequence numbers to ensure the fresh routes and guarantees loop freedom. In the Ad Hoc networks, the route changes frequently because of the mobility of nodes, as a result, the routes maintained by certain nodes, become invalid. The sequence numbers make it possible to use the fresh routes. In AODV, the route table is used to store routing information. Additionally, node maintains a list of precursor nodes to route through them to reach the destination. A source node broadcasts a route request (RREQ) packet on the network each time the route to the destination is not available or the previously entered route is inactivated.

Each middle node that receives the RREQ checks its own routing table, if it is the destination, it sends backs an RREP (Route Reply) packet. If not, it relays the RREQ packet by broadcasting it to its neighbors. In the absence of response RREP, this process repeated until destination node or intermediate node that has a fresh enough route toward destination node receives RREQ, and in this way, it creates RREP message, and inversely sends along the reverse route established at intermediate nodes during the route discovery process to source node. The number of the RREQ message steps increases by passing through each node. Nodes sending RREP updates the sequence number for the source node in its own route tables. The source receives a RREP when the sequence number of RREQ is smaller than the sequence number of its routing table, or the same sequence number with a smaller hop count, the newer RREQ is removed. If there are two routes toward receiving destination, then the node selects the route with maximum sequence number or if sequence numbers are the same, the message with minimum number of steps is selected [23]. Sequence number acts as a time stamp. By using sequence numbers, nodes can recognize those that sent and transferred information, and which node is newer than the others [24]. The intermediate nodes and the source node store the next hop information. The routing table entry contains the following information:

Destination node, Destination sequence number, Next hop, Number of hops, Hop Count, Active neighbours for the route, and expiration timer.

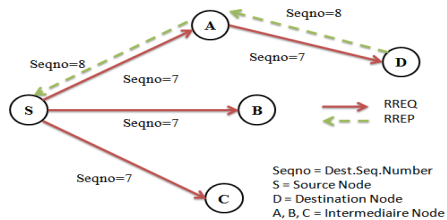

Figure2.1: AODV route discovery process

From the figure 2.1, the source node S broadcast the RREQ to reachable neighbors A, B and C to find the best possible route to the destination node D. After receiving the RREQ, node A, B and C either:

- Send out the RREP message to the source node if it is the destination or it is an intermediate node with a fresh enough route to the destination with a higher sequence number or equal to  the RREQ message,
- Update the routing table and broadcasting the RREQ until the destination node or intermediate node with fresh enough route.

The destination node D receiving the RREQ message from node A and forwards the RREP message to this node. Node A sends the RREP message to node S and updates its routing table. Source node S also update the routing table for the new route to the destination node D using the AODV recvReply() function. The explanation of normal recvReply() mechanism is described as follows:

---

**Algorithm 1** Algorithm of AODV recvReply() function

---

RecvReply (Packet P)
If (P.dst no entry in Routing Table RT) Then
Add entry of P.dst to RT
End
Select dst_seqno from RT
If (P.dst_seqno>RT.dst_seqno or P.dst_seqno=RT.dst_seqno and P.hops < RT.hops) Then
Update RT entry with P
Send data packets to the route in RT
Else If (routing is UP for P) Then
Forward packet P
Else discards P
End End If
End

---

## 2.2 BLACK HOLE ATTACK

Routing protocols are having a variety of attacks. In which a malicious node sends forged RREP packet to inform the nodes of that it has the shortest path to the node whose packets it wants to intercept, we talk about black hole attack [20]. In other words, a malicious node uses the routing protocol (such as AODV) to promote false information of having shortest path to the destination node or to the packet it wants to intercept, then black hole will  have the accessibility in replying to the route request and creates a reply where an extremely short route is advertised. If the malicious reply reaches node before the reply from the actual node, a forged route created. When the attacker inserts itself between the communicating nodes, it is able to drag the packets towards them [3]. And when the source receives these false RREP, it starts transmitting the data packets to the black hole node instead of transmitting them to the destination.

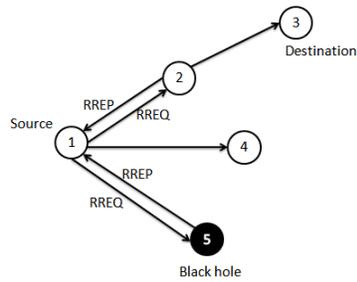

Figure2.2: Routing discovery in AODV with black hole attack

For example in the figure 2.2, the source node (node 1) broadcasts a route request packet RREQ to its neighbours to find a route to the destination node (node 3). It is assumed that routing table of the intermediate node 2 has a route to the destination node and node 5 is a malicious one in the network. So the node 5 automatically sends a false RREP to (node 1) without checking the routing table. The malicious RREP attained rapidly to the node 1 before the responses of other nodes in the network. Now, node 1 accepts the shortest route through the node 5 and sends application layer data to the node 3 via this node rejecting other RREP packets (in this case, a RREP packet from node 2). The black hole node drops all data packets rather than forwarding them to the destination, unfortunately, source node sends the data packets assuming that this data would reach safely the destination node. The intention of implementing a black hole in the network may be as simple as disrupting the normal network operation to as severe as man in the middle attack or denial of service attack. The following code will be added in the AODV protocol to generate the false answers by the black hole attacker: recvRequest(Packet*) function.

---

**Algorithm 2** Extract of black hole algorithm

if
```
 (rt && blackhole == 1) Then
 assert (rq- > rq_dst == rt- >  rt_dst);
 sendReverse (rq- > rq_src); // IP Destination
 rq- > rq_timestamp; // timestamp
 rt- > pc_insert (rt0- > rt_nexthop);
 rt0- > pc_insert(rt- > rt_nexthop);
 Packet::free(p);
End If
```

---

## 3. AODV CONTROL

Security in Ad Hoc networks is an essential component, especially for those security-sensitive applications. We analyze the security in Ad Hoc based on the following attributes: Integrity to guarantee that the messages of routing exchanged between the entities were never corrupted. Authentication to verify the identity of an entity or a node in the network and the non-repudiation to verify that the sender and recipient are parties that they say have respectively sent or received the message [17]. In the case of the AODV routing protocol, many solutions have been explored. Our proposed approach will be developed basing on these solutions and using the concept of fidelity levels.

### 3.1 RELATED WORKS

Security solutions are always been one of the most essential important issue in the Ad Hoc network communication. Various methods have been developed to protect the network from black hole with the AODV protocol. We will discuss several security solutions offered by researchers for this kind of attack.

The ARIADNE protocol [8] offers a solution that provides authentication point-to-point of routing messages using secret key hash functions (HMAC: Hash-based Message Authentication Code). However, to ensure a secure authentication, ARIADNE is based on TESLA [15] which is a protocol that ensures safe authentication during broadcasts. Each node has a secret key that allows it to calculate a hash chain which is subsequently used as follows: when it transmits a route request message, the node adds an HMAC calculated on the entire message with the last hash still not used in the chain. If a route replayed to a particular node, this one reveals the value used in the route request. When all nodes on the way perform this operation, the path is authenticated. To function, ARIADNE requires that all nodes in the network are synchronized (through use of TESLA) and that each one of them knows the last value of the hash function of all others. This extension makes it possible to make safe the protocol against the attacks by modification and manufacturing but is still vulnerable to egoistic behavior.

Xiaoqi et al. [21] gave a TAODV (Trusted Ad Hoc on Demand Routing Protocol) to secure AODV, it's a routing protocol based on applying trust model to secure mobile Ad Hoc network. TAODV has several salient features like: the trust and trust relationship among nodes that can be represented, calculated and combined for efficient routing; a malicious node will eventually be detected and denied to the network and the performance of the system is improved by avoiding requesting and verifying certificates at every routing step.

Sanzgiri et al. [19] designed a secure AODV algorithm which called ARAN. In this protocol, nodes use public key certificates to authenticate themselves to other nodes during the routing process. ARAN relies on the use of authentication, non-repudiation and message integrity in Ad Hoc networks by using a cryptographic certificate which is followed by a route instantiation process that ensures end-to-end security services. The main disadvantage that ARAN uses the trusted certification server and its requires every node to sign the message before transmitting, which is very costly in terms of power and the size of the routing messages increase at each hop.

Another example of protocol based on the reputation was presented by Buchegger and Boudec [5]. CONFIDENT (Cooperation of Node Fairness in Dynamic Ad-hoc Network) interacts with the misbehaving node in the network. The reputation system based CONFIDANT scheme punishes the misbehaving nodes by the detection and the isolation from the network, but if the nodes use limited transmission power this protocol cannot work correctly in network.

Zapata have proposed the SAODV [22] as an extension of the AODV, that can be used to protect the routing messages (RREQs, RREPs, and RERRs), based on public key cryptography. SAODV use two security mechanisms: digital signatures and hash chains. The first are used to authenticate RREQ and RREP messages and the second are used to authenticate the hop count, the only changeable information in the packets. A hash chain is formed by repeatedly applying a one way hash function to a seed. SAODV requires the existence of suitable asymmetric crypto system, where each node has a pair signature key. Furthermore, each node is capable to acquire and verify the association between the address and the public key of other nodes that participate in the Ad Hoc network. However, the use of hash chains does not make it possible to prevent all the attacks on the number of hops. Also, although the hash hop count prevents a prospective malicious node to announce shorter routes that in reality, nothing prevents an attacker to increase arbitrarily the length of the routes. Indeed, such a node can apply the hash function several consecutive times before relaying a packet, and then the route appears longer than it is in reality [22]. Thus, SAODV has disadvantages, nothing prevents a node from leaving a hop count unchanged or increasing it arbitrarily. Malicious node can acquire routes by consistently declaring high hop counts and it can impersonate another node while forwarding a fake RREP. In addition, in the event that there are several attackers accomplices, an attack of type black hole can always be launched and the number of hops can even be decremented on arrival, in a transparent way for the other nodes. Hence, encryption solution approaches do not address packet dropping by a black hole node.

A mechanism has been proposed by Vishnu K et al. [13], it is capable to detect and remove the collaborative malicious nodes which introduce huge packet drop from network. A purpose of Backbone network consists of group of strong nodes in terms of battery power and these nodes can be allowed to allocate the RIP to the newly arrived nodes. Before transmitting data packets, the source node asks the backbone network to allocate RIP address. When the backbone network assigns the RIP address, the source node sends RREQ to search for destination and also for allocated RIP. If the source node only receives the RREP from the destination then network is safe, but if RREP comes from RIP then it implies that adversary might be existed in the network. As a result, the source node sends a monitor message to alert these neighborhoods. Then the neighbor nodes broadcast this alert message through the whole network and it sends a reply message to the source node that there is black hole node in the network.

## 3.2 PROPOSED APPROACH

### 3.2.1 PROBLEM STATEMENT

The vulnerability of an Ad Hoc may take place once the latter is the subject of an attack. Malicious nodes can perform many types of attacks for the dysfunction of the network, especially at the time of the routing between the mobile equipments. In our case, the routing is supposed to be provided by the AODV protocol which is subject to a black hole attack. The presence of this kind of attack degrades the performances in terms of security and efficiency.

In order to improve the prevention of this protocol against the black hole attack, our approach consists in a coupling of security mechanisms used in the SAODV protocol and the concept of fidelity which is closely related to the nodes in the Ad Hoc network. The resulting algorithm uses the cryptography with public key through the digital signatures and hash functions, with the addition of a fidelity table where in every participating node will be attributed a fidelity level that acts as a measure of trustworthiness of that node. With on this basis, the PC-AODV-BH protocol is proposed to select and maintain the safer routes for the routing data while neutralizing the effect of malicious nodes trying to cooperate with in the network to disturb its operation. The contribution of PC-AODV-BH protocol to improve the security of AODV is a simple idea considering the consistency of secure AODV protocol (SAODV) in terms of cryptographic tools, but their connection with the fidelity of nodes has opened other ways such as the comparison between the two protocols in terms of efficiency facing the black hole disturbance.

### 3.2.2 MECHANISMS OF SAODV PROTOCOL

In order to secure the AODV protocol, Zapata [22] conceived the SAODV protocol, a secure variant based on digital signatures and hash functions.

The role of a hash chains is to keep the integrity of the hop count which is supposed to be incremented at each hop. In such a way that enables the intermediate or the destination node that receives the messages to verify that the hop count has not been decremented by a malicious node [22]. This function is widely used in cryptography, in order to reduce the size of data to be processed by the encryption function. Indeed, the main feature of a hash function to produce a hash data is to say a condensate of these data. This digest has a fixed size and which the value differs depending on the function. Among the usual functions, we can cite (MD4, MD5, SHA-1 and SHA-2) [6].

When in their turn, digital signatures authenticate the non-mutable data in the RREQ and RREP packets. That means that they sign everything but not the hop count of the AODV messages and the Hash from the SAODV extension. The final destination node signs the RREP generated by an intermediate, so that is the main problem in the application of the digital signatures. To remedy this problem, SAODV has two types of signatures as the single (SS) and double (DS) signature [22]. The SS is used to send a RREP request of the destination and the DS is used for sending the route response from the intermediate nodes, if it has enough new roads. Single signature is applicable in the route discovery mechanism because it is difficult to have enough paths by intermediate nodes. If disturbances occurred during the data transmission process, the source node resets the route discovery process. Moreover, the time to generate the RREP be it from the destination or from the intermediate nodes, and the signing processes are applicable and also the treatment steps of SAODV applied by the intermediate and the destination nodes.

### 3.2.3 FIDELITY

In each node in the network, the fidelity is basically considered as an integer number or a counter that is associated with it. This concept contributes to maintain the security of the network while measuring what one calls the fidelity levels [12, 18]. In other words, when the data packets are forwarded successfully, this counter is increased. According to the loyal participation of nodes in the network, their fidelity levels are updated. After successful reception of the packets by the destination node, this latter replies by sending an acknowledgement packet to the source. Due to this acknowledgement the intermediate nodes fidelity level will be incremented and the packet is exchanged. If no acknowledgement is received by the source node within a timer event, the intermediate node level will be decremented and also of the next hop of the intermediate node. The fidelity tables are exchanged periodically between the participating nodes in the network.

In the arrival of positive ACK, the source node increment the fidelity levels of the corresponding nodes and the fidelity values are exchanged (figure 3.1).

In the literature, there exist various definitions of this concept. In our work we characterize the latter by the fidelity level of each intermediate node $i$. It is the degree of participation in the network operations. In other words; it is through the reports of transfer and reception of each node. Thus, the fidelity level $\varphi_i$ of the node $i$ is given by:

$$\varphi_i = \left[\frac{(MT)_i}{(MR)_i}\right]$$

Where MT (resp. MR) is the number of forward (resp. received) messages by the node *i* and [X] indicate the integer part of the real X.

Whenever we observe that the fidelity level value of a particular node is greater than that of another node then we can conclude that the one having the greater value is more durable than the other from whose value is greater. Because of a node with greater value signify that it is an experienced node in the network and it has exchanged packets most correctly than other. When the level of a node drops to 0 [12], it implies there is no packets faithfully forwarding and the node is considered to be a malicious node and it is eliminated from the network. When black hole is detected it should to be declared to the other nodes in the network. This is accomplished by sending alarm packets. When a node receives an alarm packet passed to the entire network, so it can identify and eliminate the use of the black hole. Figure 3.2 shows the behavioral process where the black holes working as a team have been eliminated from the whole network.

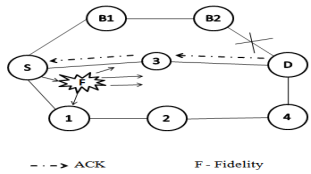

Figure3.1: Receiving acknowledgement and
broadcasting fidelity packets

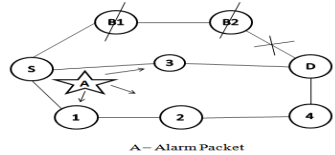

Figure3.2: Black hole nodes elimination

### 3.2.3 PC-AODV-BH ALGORITHM

To summarize the followed procedures in our approach, we present below an extract of the resulting algorithm. It is subdivided into three principal steps according to the specification of the source, intermediate and destination nodes.

---

**Algorithm 3** PC-AODV-BH Algorithm

---

*Step1 (Source node): Diffusion of the RREQ packet.*

  *1. The hash function generated ← 0*

  *2. RREQ secure ← Original RREQ + hash chain   protection + digital signature + protection key of the node*

*Step2 (Intermediate node): Receiving RREQ Packets.*

*Int   $\varphi = \left[\frac{mt}{mr}\right]$, $\varphi^c$*

*While (destination) Do*

*Receiving secure RREQ*

*If (INaddr == RREQ.DESTaddr) Then*

*Send the RREP   RREP + Hash Chain protection + digital signature + protection key*

*EndIF*

*Calculate AVG_φ_LEVEL = φ_IN + φ_nexthop*

*If (φ_IN > $\varphi^c$ and φ_nexthop > $\varphi^c$) Then*

*Send the data*

*EndIf*

*Repeat until a maximum Time to live value.*
*End*
<u>*Step3*</u> *(Destination node): Receiving the secure RREP packets.*
*Time = current time value+ timer value*
*While (current time value ≤ time) do*
*If (ACK is received) then*
*Increment φ_IN and φ_nexthop*
*Else decrement φ_IN et φ_nexthop*
*If (Fidelity level of a node = 0)) then*
*Remove the node from neighbour table and fidelity table.*
*An alarm packet is generated and diffused to all neighboring nodes concerning the malicious node.*

The source node forwards a secure and protected RREQ with a digital signature towards its neighbours. As soon as the intermediate nodes receive it this request, if one of them is the destination node, then it generates a uncast RREP encrypts and send back in the reverse route. Else if the node is the intermediate one, the calculation of the fidelity level for each node will start to select the best among them. A comparison between the average of fidelity level of the node of the current level and that in the next hop with a threshold value noted $\varphi^c$, makes it possible to authorize or prevent the transfer of data. If the average value of the fidelity level is higher than the $\varphi^c$ value, there will automatically be a sending of data packets. On receiving the data packets by the destination node, this latter will send a positive acknowledgement to the source, by which the intermediate node's level will be incremented. If within a specified time interval, if the source node doesn't receive the positive acknowledgement it will decrement the intermediate node's level to identify the cooperative black holes. In the case where the fidelity level drops to 0, it implies that there is no packets faithfully forwarding and the node is considered to be a black hole node and it is removed from the routing and fidelity tables and not only it is eliminated from the network.

## 4. SIMULATION

**Context:** To analyze the behavior of Ad Hoc routing protocols, for our simulations based on NS2 simulator, we use a CBR (Constant Bit Rate) application. All the data packets CBR are generated between nodes using a traffic generator which creates randomly CBR connections that start at moments uniformly distributed between 0 and 60 seconds with a pause time equal to 10 seconds. The size of data is 512 bytes. Mobility scenarios are generated using a random way point model (RWP) by varying the mobile nodes moving in an area of 500m x 500m. The number of nodes is fixed in 35.

Table 1 – Simulation parameters

| Parameter | Value |
|---|---|
| Simulator | NS2 |
| Number of Nodes | 35 nodes |
| Traffic Type | Constant Bit Rate CBR |
| Mobility Model Random | Waypoint |
| Terrain area | 500 m*500 m |
| Simulation Time | 60 seconds |
| Packet Size | 512 bytes |
| Routing Protocols | AODV, SAODV, PC-AODV-BH |
| Pause time | 10 seconds |

In table 1 we summarize the simulation parameters of different protocols used in this work. In order to evaluate the performance of concerned routing protocols, the following three metrics are considered:

(i) End-to-End Delay: this is the average delay between the sending of data packets by the source and the successfully receiving it by the destination.

(ii) Throughput: this is the amount of transmitted information by a communication channel according to a given time interval.

(iii) Packet loss: this metric informs us about the amount between the generated and received packets during the time of communication.

The variation of these three metrics is given according to time, and they are related to the autonomous, disturbed and controlled cases of the AODV routing protocol state.

**Autonomous system:** From the table 1, the performance evaluation of the network according to the three metric ones above, while using the routing protocol AODV, gave the following results (figures 4.1, 4.2 and 4.3).

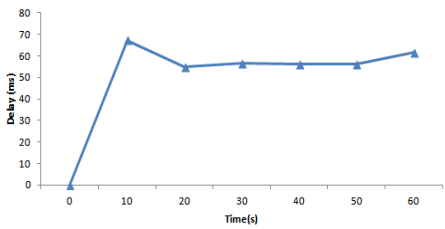
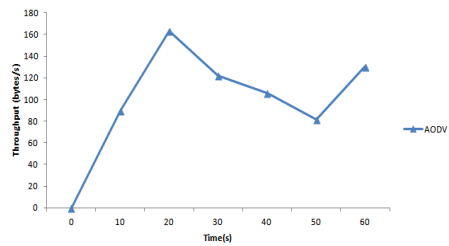

Figure4.1: End to end delay        Figure4.2: Throughput of communication

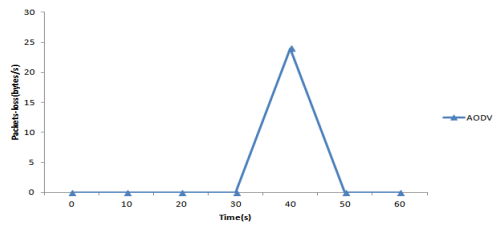

Figure4.3: Packets loss

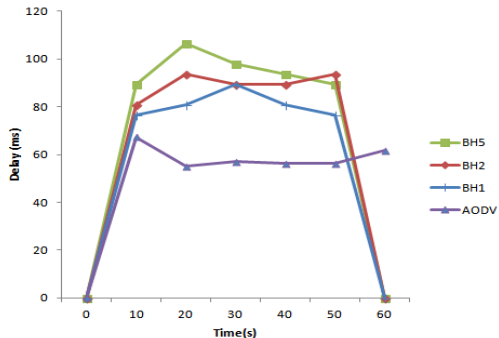
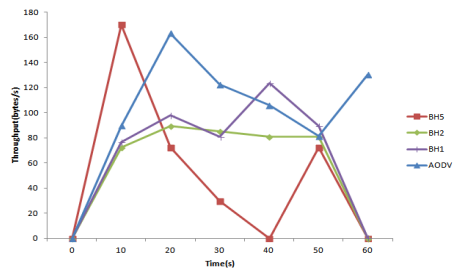

Figure4.4: End to end delay in the presence of presence of attack

Figure4.5: Throughput of communication in   the the presence of attack

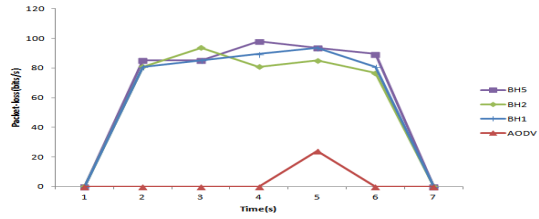

Figure4.6: Packets loss in the presence of attack

**Controlled system:** When the network described in table 1 undergoes a disturbance of the type 1, 2 or 5 black holes, our objective is to be able to cancel their effects using SAODV and PC-AODV- BH protocols, and then to compare the two controls applied in terms of effectiveness of safety of AODV routing deal with such threats. Then, an implementation of the two protocols enables us to obtain the figures 4.7 - 4.15.

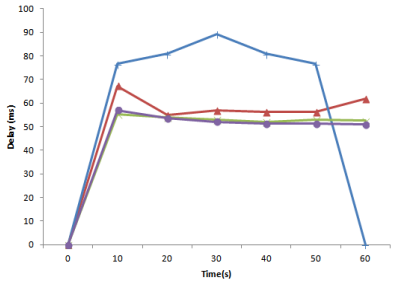

Figure4.7: Comparison of the end to end delays oneblack hole

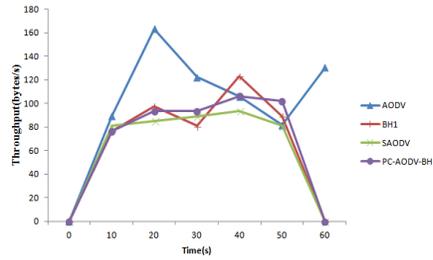

Figure4.8: Comparison of the throughputs with of communication with one black hole

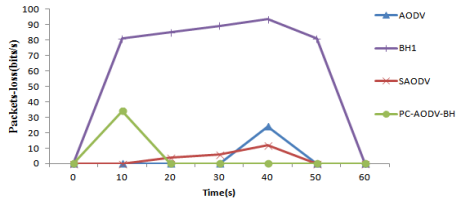

Figure4.9: Comparison of the packets loss with one black hole

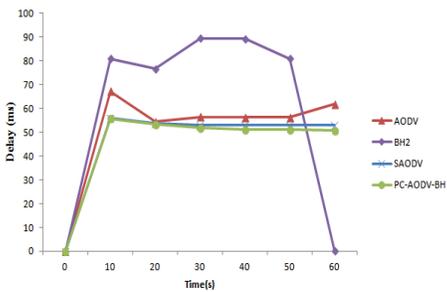

Figure4.10: Comparison of the end to end delays with two black holes

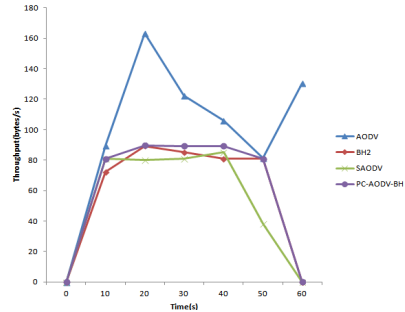

Figure4.11: Comparison of the throughputs of communication with two black holes

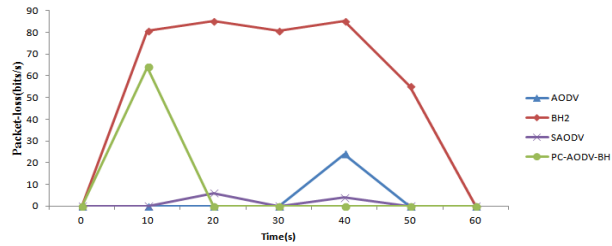

Figure4.12: Comparison of the packets loss with two black holes

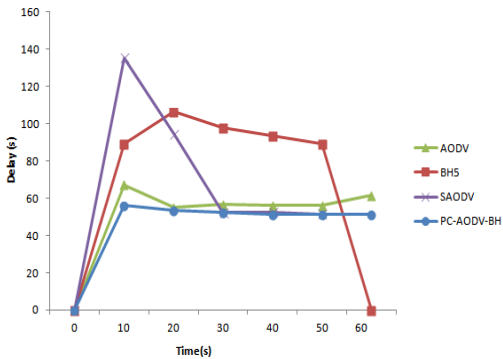

Figure4.13: Comparison of the end to end delays
with five black holes

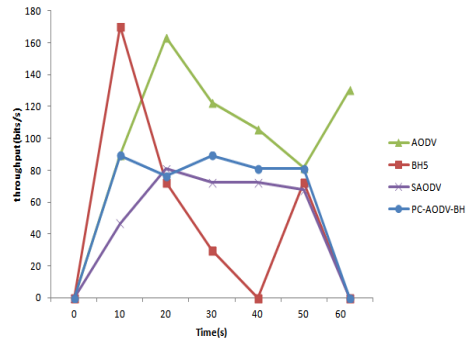

Figure4.14: Comparison of the throughputs
of communication with five black holes

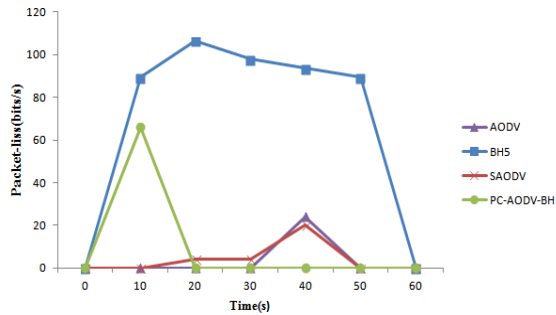

Figure4.15: Comparison of the packets loss with five black holes

**Discussion results:** The figures 4.1, 4.2 and 4.3 represent respectively the variation of the end-to-end delay, the throughput of communication and the quantity of packages lost in the AODV routing according to the time. It is about a normal evolution of this protocol, because what concerns us is the control of this protocol facing the black hole attacks.

In the presence of this kind of attack, the behavior of AODV is modified. Indeed, the figure 4.4 shows that the end-to-end delay is proportional to the number of black hole attacks over the entire interval of time. As this metric is a major challenge which any Ad Hoc network seeks to minimize, its increase based on malicious nodes is due to the cooperation of the latter for the degradation of the receptions in the entire network.

In a similar way, the increase in the quantity of the packets loss (figure 4.6) is mainly the result of the black hole attack which consists in falsifying the borrowed routes and the cooperation of the attackers as well. The figure 4.5 shows the effect of black holes attacks on the throughput of communication in the network.

The observed decrease, compared to undisturbed AODV, is due to the fact that the bandwidth is also shared by malicious nodes that cooperate and contribute to the transfer and the reception of the data by emulating the source by erroneous information.

The objective of our approach is to be able to make the disturbed system, which is in other words the Ad Hoc network in its autonomous state by using a protector control. From the 9 figures 4.7 - 4.15, we observe that the curves obtained by the implementation of SAODV and PC-AODV-BH protocols are approximately close to those corresponding to the case of the routing AODV in the autonomous case. In other words, these two protocols not only made it possible to decrease the end to end delay and packets loss, but increased the throughput of communication, as well that is caused by different malicious nodes and their cooperation. It should be noted that, even if there is not a remarkable increase in the throughput, the two protocols remain effective for the protection of the AODV protocol against black hole attack, since malicious nodes share the channel or bandwidth with other nodes in the network.

That being, one notes that the two protocols of control are able to defend the AODV routing in the three situations of attacks (1, 2 and 5 black holes), for the considered metric. Moreover, the figures 4.7 - 4.15 show that our proposed protocol PC-AODV-BH is more efficient than the SAODV protocol. That is a consequence of the use of the fidelity levels for choosing the safest route taking into account at the same time the current level and that of the next hop.

## 5. CONCLUSION

In this work, we tackled a control problem of the AODV routing protocol in the presence of   the black hole attacks in the mobile Ad Hoc networks. The attackers can easily be deployed within the network to disturb its operation. We suggested a realizable solution to control the AODV protocol to this type of attack: it is the PC-AODV-BH protocol. This is a combination of public key cryptography mechanisms and the concept of fidelity levels which are associated to each Ad Hoc nodes. The implementation of the resulting algorithm, according to the throughput of communication, packets loss and end to end delay, has shown that the proposed protocol made it possible to answer the objective of this work. On the one hand, the PC-AODV-BH was able to neutralize approximately the effects of cooperation black holes. Besides, it is considered as an improvement of SAODV protocol owing to the fact that it treats at the same time the current level of node and that of the next hop through the injection of the fidelity.

In the future, we plan to extend and develop the security mechanisms for other Ad Hoc routing protocols performance simulation and also for the delay tolerant networks (DTN) as in the case for example of the works being developed in [1, 2].

## REFERENCES

[1]   E. A. Abdellaoui Alaoui, S. Agoujil, M. Hajar and Y. Qaraai. The Performance of DTN Routing Protocols: A Comparative Study. WSEAS TRANSACTIONS on COMMUNICATIONS, volume 14, pp. 121-130, 2015.

[2]   E. A. Abdellaoui Alaoui, S. Agoujil, M. Hajar and Y. Qaraai. DTN Routing Hierarchical Topology (DRHT): Optimal Cluster Head. Frontiers of Information Technology and Electronic   Engineering, 2016. Submited.

[3]   S. K. Bobby. Simulation and Analysis of Blackhole Attack in MANETs for Performance Evaluation. International Journal of Latest Trends in Engineering and Technology, volume 2, Issue   1, pp. 186-192, 2013.

[4]   A. Boukerche, B. Turgut, N. Aydin, M.Z. Ahmad, L. Boloni and D. Turgut. Routing protocols in ad hoc networks: A survey. Journal of Computer Networks, Volume 55, Issue 13, pp.3032–3080, 2011.

[5]   S. Buchegger and J. Y. Le Boudec. Performance analysis of the CONFIDANT protocol. Proceedings

of the 3rd ACM international symposium on mobile ad hoc networking and computing, 2002.

[6]     R. Glabb, L. Imbert, G. Jullien, A. Tisserand and N. Veyrat-Charvillon. Multi-mode operator  for SHA-2 hash functions. Journal of Systems Architecture 53, pp. 127–138, 2007.

[7]     H. Deng, W. Li, D.P. Agrawal. Routing Security in Wireless Ad Hoc Networks. IEEE Communications Magazine, Volume 40, Octobre 2002.

[8]     Y. Hu, A. Perrig and D. Johnson. Ariadne:  A Secure On-Demand Routing Protocol for Ad  Hoc Networks. Wireless Networks, Volume 11, pp. 21-38, 2005.

[9]     Jasvinder, Monika Sachdeva. A Survey of Behavior of MANET Routing Protocols Under Black-hole Attack. International Journal of Advanced Research in Computer Science and Software Engineering, Volume 3, Issue 8, pp. 647-651, 2013.

[10]   Huaizhi Li, Zhenliu Chen and Xiangyang Qin. Secure Routing in Wired Networks and Wireless Ad Hoc Networks. IEEE, 2004.

[11]   M. Shahraeini, M. H. Javidi and M. S. Ghazizadeh. Comparison between Communication Infrastructures of Centralized and Decentralized Wide Area Measurement Systems. IEEE TRANSACTIONS ON SMART GRID, VOL. 2, NO. 1, pp. 206-211, 2011.

[12]   Nikos Komninos, Dimitris Vergados and Christos Douligeris. Layered security design for mobile ad hoc networks. Journal of computer and security, 25, pp. 121-130, 2006.

[13]   Amos J Paul and Vishnu K. Detection and Removal of Cooperative Black/Gray hole attack in Mobile Ad hoc Networks. International Journal of Computer Applications (ISSN NO. 0975- 8887), Volume 1, Issue 22, 2010.

[14]   C.E. Perkins and E.M. Royer. Ad-hoc on-demand distance vector routing. Proceedings of the 2nd IEEE Workshop on Mobile Computing Systems and Applications, pp. 90-100, 1999.

[15]   A. Perrig, R. Canetti, J. D. Tygar and D. Song. The TESLA broadcast authentication protocol. RSA Crypto Bytes, Volume 5, Issue 2, pp. 2-13, 2002.

[16]   Y. Qaraai and A. S. Bernoussi. Protector control: Extension to a class of nonlinear distributed system. Int. J. of Appl. Math. Comput. Sci., Volume 20, Issue 3, pp. 427-443, 2010.

[17]   A. RACHEDI. Contribution à la sécurité dans les réseaux mobiles ad Hoc. Université d'Avignon et des Pays de Vaucluse, 2008.

[18]   Himadri Nath Saha, Debika Bhattacharyya and P.K. Banerjee. Fidelity Based On Demand Secure (FBOD) Routing in Mobile Adhoc Network. International Journal of Advanced Com- puter Science and Applications, pp. 615-627, 2011.

[19]   K. Sanzgiri and B. Dahill and B.N.Levine and C.Shields and E.M. Belding-Royer. A Secure Routing Protocol for Ad hoc Networks. In proc. Of International Conference on Network protocols (INCP), 2002.

[20]   S. Sharma and R. Gupta. Simulation study of blackhole attack in the mobile ad-hoc networks. Journal of Engineering Science and Technology, Volume 4, Issue 2, pp. 243-250, 2009.

[21]   Xiaoqi Li, Michael R. Lyu and Jiangchuan Liu. A Trust Model Based Routing Protocol for Secure Ad Hoc Networks. Aerospace Conference, IEEE, 2004.

[22]   M. Guerrero Zapata and N. Asokan. Securing Ad hoc Routing Protocols, Proceedings of the 2002 ACM Workshop on Wireless Security (WiSe 2002), pp. 1-10, 2002.

[23]   I. Zangeneh, S. Navaezadeh and A. Jafari. Investigating the Effect of Black Hole Attack on AODV and DSR routing protocols in Wireless Ad Hoc network. Journal of Advances in Com- puter Research, Volume 5, Issue 1, pp. 13-20, 2014.

[24]   Lidong Zhou and Zygmunt J. Haas. Securing Ad Hoc Networks. Proceedings of IEEE Net- work and Computer Communications, pp. 24-30, 2011.